\documentclass[12pt,a4paper,twoside,english]{article}

\usepackage{babel}

\begin{document}

\title{Quasinormal frequencies of $D$-dimensional Schwarzschild black holes: evaluation via continued fraction method.}

\author{Andrzej Rostworowski}
\date{}
\maketitle
\begin{center}
M. Smoluchowski Institute of Physics, Jagellonian University, Reymonta 4, 30-059 Krak\'ow, Poland
\\
arostwor@th.if.uj.edu.pl
\end{center}

\abstract{We adopt Leaver's \cite{Leaver} method to determine quasi normal frequencies of the Schwarzschild black hole in higher ($D \geq 10$) dimensions. In $D$-dimensional Schwarzschild metric, when $D$ increases, more and more singularities, spaced uniformly on the unit circle $|r|=1$, approach the horizon at $r=r_h=1$. Thus, a solution satisfying the outgoing wave boundary condition at the horizon must be continued to some mid point and only then the continued fraction condition can be applied. This prescription is general and applies to all cases for which, due to regular singularities on the way from the point of interest to the irregular singularity, Leaver's method in its original setting breaks down. We illustrate the method calculating gravitational vector and tensor quasinormal frequencies of the Schwarzschild black hole in $D=11$ and $D=10$ dimensions.\\
We also give the details for the $D=9$ case, considered in \cite{bcrst}.}

\section{Introduction and setup.}
\noindent 
Our motivation to study quasinormal modes of Schwarzschild black holes in higher dimensions comes mainly from the possibility of studying the dynamics of gravitational collapse in vacuum initiated with the work of Bizo\'n, Chmaj and Schmidt \cite{bcs}. At the expense of going to higher ($D \geq 5$) odd dimensions, they showed how to evade Birkhoff's theorem and study gravitational collapse in vacuum at radial symmetry. It was shown \cite{bcs, bcrst} that for $D=5, \, 9$, the D-dimensional Schwarzschild black hole is the attractor for large initial data and at some intermediate times the solution settling down to the Schwarzschild black hole, obtained from nonlinear numerical evolution, is well approximated outside the horizon by the least damped quasinormal mode. Therefore the precise values of fundamental quasinormal frequencies of Schwarzschild black holes in odd dimensions are urgently needed, as they help to check the validity of the numerical code used in evolution. The reliable values of quasinormal frequencies of Schwarzschild black hole are available for $D=5$ case \cite{cly1, cly2}, but for $D>5$ only results from WKB methods are published \cite{k1, k2, bcg} and it is known that they may be not accurate for small values of angular momentum and/or higher overtones. Therefore getting these values with Leaver's method of continued fractions \cite{Leaver}, giving the most precise values of quasinormal frequencies in $D=4$ and $D=5$ dimensions, is worthwhile. We describe below how to modify Leaver's method to obtain gravitational vector and tensor quasinormal frequencies of Schwarzschild black hole in $D \geq 10$ dimensions.
\\   
The line element of the Schwarzschild solution in D-dimensions has the form
\begin{equation}
\label{SchwarzschildD}
ds^2=A(r)dt^2 - A^{-1}(r)dr^2 - r^2\,d\Omega^2_{D-2},
\end{equation}
with 
\begin{equation}
\label{laps}
A(r)=1-\left(\frac {r_h} {r} \right)^{D-3}.
\end{equation}
In what follows we take $r_h=1$. In linear approximation the radial component of gravitational vector and tensor  perturbation of the metric (\ref{SchwarzschildD}) satisfies the following Schr\"odinger type differential equation
\begin{equation}
\label{master}
\frac {d^2} {dx^2} \psi + A(r) \left( \frac {L(L+D-3)} {r^2} + \frac {(D-2)(D-4)} {4r^2} + \frac {(1-s^2)(D-2)^2} {4r^{D-1}} \right)\psi = k^2 \psi,
\end{equation}
where the tortoise coordinate $x$ is defined by $dx/dr\,=\,A^{-1}(r)$ and the parameter $s$ depends on the type of the perturbation ($s=0$ for the gravitational tensor and $s=2$ for the gravitational vector perturbation). Eq. (\ref{master}), derived independently by Gibbons and Hartnoll \cite{gh} and Ishibashi and Kodama \cite{ik}, generalises the well known Regge-Wheeler equation \cite{rw} to D-dimensions. 
In this setting quasi normal modes are defined as solutions of (\ref{master}), satisfying the outgoing wave boundary conditions
\begin{equation}
\label{bc}
\psi \stackrel{x \rightarrow \pm \infty}{\sim} \exp (\pm i k x),
\end{equation}
with Im$(k)<0$. The corresponding values of $k$ are called quasi normal frequencies. Eq. (\ref{master}) has $D-2$ regular singular points (at $r=0$ and at $D-3$ roots of $r^{D-3}=1$) and the irregular singular point at infinity. 
Leaver's method \cite{Leaver} of determining quasi normal frequencies consists in separating boundary behavior and then transforming $r$ into $\rho(r)$ in such a way that the singularities at $r=1$ (horizon) and at $r=\infty$ become the closest singularities in the $\rho$ plane. In $D=4$ it is accomplished with the substitution
\begin{equation}
\label{LeaverD4}
\psi(r) = (r-1)^{-ik}r^{2ik}e^{ikr} u\left( \rho(r) \right),
\end{equation}
where
\begin{equation}
\label{series}
u\left( \rho(r) \right) = \sum _{n=0}^{\infty} a_n \rho^n = \sum _{n=0}^{\infty} a_n \left(\frac {r-1} {r} \right)^n,
\end{equation}
and the coefficients $a_n$ are given by the 3-term reccurence relation
\begin{equation}
\label{reccurenceD3}
\gamma^{(1)}_n a_{n+1} + \gamma^{(2)}_n a_{n} + \gamma^{(3)}_n a_{n-1} = 0,
\end{equation}
with the initial condition $a_{0}=1$, $a_{-1}=0$ and $\gamma^{(j)}_n$  given in \cite{Leaver}. Then, the quantization condition is the convergence of the series (\ref{series}) on the convergence radius $\rho=1$. The two linearly independent solutions of the reccurence behave as 
\begin{equation}
\label{anlimit}
a_n \stackrel{n \rightarrow \infty}{\sim} \exp \left( \pm \sqrt{-8 i k n} \right),
\end{equation}
thus the minimal solution makes the series (\ref{series}) convergent at $\rho=1$. The discrete values of $k$ for which the solution given by the initial condition $a_{0}=1$ and $a_{-1}=0$ is a minimal one define quasi normal frequencies. For these values the following equation, involving an infinite continues fraction holds
\begin{equation}
\label{continuedfrac}
\frac{a_1}{a_0} = \frac{\gamma^{(2)}_0}{\gamma^{(1)}_0 } = - \frac {\gamma^{(3)}_1} {\gamma^{(2)}_1 -} \, \frac {\gamma^{(1)}_1 \gamma^{(3)}_2} {\gamma^{(2)}_2 -} \, \frac {\gamma^{(1)}_2 \gamma^{(3)}_3} {\gamma^{(2)}_3 -}\dots \,.
\end{equation}
To determine the quasi normal frequencies we truncate the infinite continued fraction in (\ref{continuedfrac}) at some denominator and seek for the solutions of (\ref{continuedfrac}) which are stable with respect to the change of depth of this truncation.
\\ 
In general for even $D>4$ the substitution 
\begin{equation}
\label{even}
\psi(r) = \left(\frac {r-1} {r}\right)^{-ik/(D-3)}e^{ikr} \sum _{n=0}^{\infty} a_n \left(\frac {r-1} {r} \right)^n 
\end{equation}
leads to a $(2(D-3)+1)$-term reccurence relation, while for odd $D$ the substitution 
\begin{equation}
\label{odd}
\psi(r) = \left(\frac {r-1} {r+1}\right)^{-ik/(D-3)}e^{ikr} \sum _{n=0}^{\infty} a_n \left(\frac {r-1} {r} \right)^n 
\end{equation}
leads to a $2(D-3)$-term reccurence relation. 
These reccurence relations can be reduced to 3-term ones using Gauss elimination as in \cite{Leaver2}. However, as $D$ increases, more and more of the $D-3$ singularities, spaced uniformly on the unit circle $|r|=1$, approach the horizon at $r=1$ and no simple transformation can move them away from the circle centered at the horizon  and the radius corresponding to $r=\infty$ in the $\rho$ plane. In the case of eq. (\ref{master}), this difficulty arises first at $D=10$. (The transformation $\rho(r)$ has to be simple enough to be easily inverted into $r(\rho)$, in our case it is a homography). Therefore, for $D \geq 10$ the solution starting from the horizon has to be continued through some mid points $0<\rho<1$, laying within the convergence radius of the presently used series representation of the solution, and Leaver's continued fraction condition can be applied only if it is the irregular singularity corresponding to $r=\infty$, which limits the convergence radius of the presently used series representation. 

\section{The $D=11$ case.}
As an example, to illustrate how the above prescription works, we determine quasinormal frequencies of the Schwarzschild black hole in $D=11$ dimensions. We choose $D=11$ as, due to our motivation given in the introduction, we are interested in odd dimensions and $D=11$ is the smallest odd dimension in which Leaver's method in its original setting breaks down for D-dimensional generalization of Regge-Wheeler potential (\ref{master}). In eq.(\ref{master}) we substitute
\begin{equation}
\psi(r) = \left(\frac {r-1} {r+1}\right)^{-ik/(D-3)}e^{ikr} u\left(\rho(r)\right), \qquad  \rho(r) = \frac {r-1} {r}. 
\end{equation}
The singular points of eq.(\ref{master}) at $r = 1, \, e^{\pm i \pi / 4}, \, \infty$ are transformed into $\rho= 0, \, 1-1/\sqrt{2} \pm i/\sqrt{2}, \, 1$ respectively (other singular points are placed at $|\rho|>1$). The singularities at $\rho= 1-1/\sqrt{2} \pm i/\sqrt{2}$ limit the convergence radius of the series representation of the solution $u\left( \rho \right) = \sum _{n=0}^{\infty} a_n \rho^n$, to $\sqrt{2-\sqrt{2}} \approx 0.76$. We choose $\rho_0 = 1/2$ (which is a regular point) as a mid point. We have 
\begin{equation}
u\left( \rho \right) = \sum _{n=0}^{\infty} a_n \rho^n  = \sum _{n=0}^{\infty} \tilde{a}_n (\rho - \rho_0)^n,
\end{equation}
where
\begin{equation}
\label{atildas}
\tilde{a}_0 = \sum _{n=0}^{\infty} a_n \rho_0^n, \qquad \tilde{a}_1 = \sum _{n=1}^{\infty} n a_n \rho_0^{n-1}.
\end{equation}
The coefficients $\tilde{a}_n$ fulfill $2(D-3)+1=17$-term reccurence relation, which reduced to the 3-term one via Gauss elimination \cite{Leaver2}, and then inserted into (\ref{continuedfrac}) yields quasinormal frequencies. 
\\
All reccurence relations are obtained analytically in \textit{Mathematica} computer algebra package. Then all other tasks (finding $\tilde{a}_1 / \tilde{a}_0$ from eq. (\ref{atildas}), reduction of the $(2(D-3)+1)$-term reccurence relation for $\tilde{a}_n$ to the 3-term one via Gauss elimination and finding the roots of the continued fraction relation (\ref{continuedfrac})) are done numerically, by the program in the $C$ programing language. To determine the roots of the continued fraction relation (\ref{continuedfrac}) we use Newton-Raphson root searching algorithm \cite{nr}.  
\\
The first three quasinormal frequencies for vector and tensor gravitational perturbations of the Schwarzschild black hole, for different values of $L$, are given in table \ref{D11}. Our values of fundamental frequencies for tensor modes are consistent with \cite{k1} (in \cite{k1} the values of fundamental frequencies for scalar field perturbations were given and the scalar field perturbation obeys exactly the same equation as the tensor gravitational perturbation). In order to compare the values obtained from our modification of Leaver's method with the values given in \cite{bcg} we also calculate the first three quasinormal frequencies for vector and tensor gravitational perturbation of the Schwarzschild black hole in $D=10$ dimensions. They are listed in table \ref{D10}. Comparing with \cite{bcg} we see a perfect agreement for larger values of $L$. This makes us feel cofident in our results. However for smaller $L$ values, and especially for the overtones there are differences exceeding 10\%. As Leaver's method works well both for smaller and larger values of $L$, we believe that the error is on the WKB method side (see the comments in \cite{bcg}). 
\begin{table}
\begin{center}
{\scriptsize
\begin{tabular}{|c|c|c|c|c|c|c|}
\hline
$D=11$ & \multicolumn{3}{c|}{vector modes} & \multicolumn{3}{c|}{tensor modes}\\
\hline
L & $n=0$ & $n=1$ & $n=2$ & $n=0$ & $n=1$ & $n=2$ \\ 
\hline
 2 & $3.6788 -1.0588 i$ & $2.3419 -2.8190 i$ & $1.2130 -3.9731 i$ & $4.3920 -1.0577 i$ & $3.3356 -3.0313 i$ & $1.9912 -3.8491 i$ \\
 3 & $4.4533 -1.0331 i$ & $3.4147 -2.9417 i$ & $1.9955 -3.7743 i$ & $5.1231 -1.0507 i$ & $4.2669 -3.0765 i$ & $2.7018 -4.0946 i$ \\
 4 & $5.2343 -1.0187 i$ & $4.4049 -2.9628 i$ & $2.8490 -4.0060 i$ & $5.8540 -1.0463 i$ & $5.1305 -3.0910 i$ & $3.4995 -4.4424 i$ \\
 5 & $6.0134 -1.0120 i$ & $5.3226 -2.9745 i$ & $3.7955 -4.3603 i$ & $6.5849 -1.0435 i$ & $5.9561 -3.0967 i$ & $4.4432 -4.7715 i$ \\
 6 & $6.7881 -1.0097 i$ & $6.1936 -2.9862 i$ & $4.8328 -4.6430 i$ & $7.3160 -1.0415 i$ & $6.7587 -3.0994 i$ & $5.4481 -4.9519 i$ \\
 7 & $7.5579 -1.0096 i$ & $7.0340 -2.9976 i$ & $5.8529 -4.7970 i$ & $8.0471 -1.0401 i$ & $7.5459 -3.1007 i$ & $6.4039 -5.0333 i$ \\
 8 & $8.3232 -1.0105 i$ & $7.8533 -3.0082 i$ & $6.8146 -4.8826 i$ & $8.7783 -1.0390 i$ & $8.3226 -3.1014 i$ & $7.3091 -5.0753 i$ \\
 9 & $9.0846 -1.0120 i$ & $8.6576 -3.0177 i$ & $7.7286 -4.9366 i$ & $9.5095 -1.0383 i$ & $9.0914 -3.1019 i$ & $8.1784 -5.1000 i$ \\
10 & $9.8425 -1.0136 i$ & $9.4505 -3.0262 i$ & $8.6084 -4.9742 i$ & $10.2408 -1.0377 i$ & $9.8544 -3.1021 i$ & $9.0222 -5.1159 i$ \\
11 & $10.5976 -1.0152 i$ & $10.2348 -3.0337 i$ & $9.4632 -5.0021 i$ & $10.9721 -1.0372 i$ & $10.6128 -3.1023 i$ & $9.8473 -5.1268 i$ \\
\hline
\end{tabular}
}
\end{center}
\caption{{\small The first three quasinormal frequencies for vector and tensor perturbation of the Schwarzschild black hole in $D=11$ dimensions.}}
\label{D11}
\end{table}
\begin{table}
\begin{center}
{\scriptsize
\begin{tabular}{|c|c|c|c|c|c|c|}
\hline
$D=10$ & \multicolumn{3}{c|}{vector modes} & \multicolumn{3}{c|}{tensor modes}\\
\hline
L & $n=0$ & $n=1$ & $n=2$ & $n=0$ & $n=1$ & $n=2$ \\ 
\hline
 2 & $3.2334 -0.9603 i$ & $2.0119 -2.7275 i$ & $0.9784 -3.5136 i$ & $3.9209 -0.9621 i$ & $3.0410 -2.8514 i$ & $1.5315 -3.5723 i$ \\
 3 & $3.9946 -0.9337 i$ & $3.1233 -2.7391 i$ & $1.5680 -3.5028 i$ & $4.6311 -0.9555 i$ & $3.9309 -2.8507 i$ & $2.2219 -3.9757 i$ \\
 4 & $4.7607 -0.9211 i$ & $4.0833 -2.7243 i$ & $2.4839 -3.9023 i$ & $5.3414 -0.9515 i$ & $4.7545 -2.8456 i$ & $3.1902 -4.4737 i$ \\
 5 & $5.5225 -0.9165 i$ & $4.9654 -2.7248 i$ & $3.6165 -4.2895 i$ & $6.0519 -0.9489 i$ & $5.5446 -2.8411 i$ & $4.3063 -4.6452 i$ \\
 6 & $6.2781 -0.9158 i$ & $5.8018 -2.7316 i$ & $4.7032 -4.4370 i$ & $6.7627 -0.9472 i$ & $6.3149 -2.8377 i$ & $5.2808 -4.6848 i$ \\
 7 & $7.0278 -0.9168 i$ & $6.6097 -2.7400 i$ & $5.6773 -4.5027 i$ & $7.4735 -0.9460 i$ & $7.0722 -2.8350 i$ & $6.1763 -4.6985 i$ \\
 8 & $7.7725 -0.9185 i$ & $7.3984 -2.7484 i$ & $6.5828 -4.5412 i$ & $8.1845 -0.9451 i$ & $7.8206 -2.8330 i$ & $7.0262 -4.7042 i$ \\
 9 & $8.5128 -0.9204 i$ & $8.1735 -2.7561 i$ & $7.4452 -4.5675 i$ & $8.8955 -0.9444 i$ & $8.5624 -2.8315 i$ & $7.8469 -4.7067 i$ \\
10 & $9.2495 -0.9222 i$ & $8.9385 -2.7631 i$ & $8.2785 -4.5869 i$ & $9.6066 -0.9439 i$ & $9.2994 -2.8303 i$ & $8.6472 -4.7079 i$ \\
11 & $9.9832 -0.9240 i$ & $9.6957 -2.7692 i$ & $9.0911 -4.6021 i$ & $10.3178 -0.9435 i$ & $10.0327 -2.8293 i$ & $9.4327 -4.7084 i$ \\
12 & $10.7144 -0.9256 i$ & $10.4469 -2.7745 i$ & $9.8882 -4.6143 i$ & $11.0290 -0.9432 i$ & $10.7629 -2.8285 i$ & $10.2070 -4.7085 i$ \\
13 & $11.4434 -0.9270 i$ & $11.1932 -2.7792 i$ & $10.6733 -4.6243 i$ & $11.7402 -0.9429 i$ & $11.4908 -2.8279 i$ & $10.9725 -4.7085 i$ \\
14 & $12.1707 -0.9282 i$ & $11.9354 -2.7833 i$ & $11.4490 -4.6327 i$ & $12.4514 -0.9427 i$ & $12.2166 -2.8273 i$ & $11.7310 -4.7084 i$ \\
15 & $12.8963 -0.9293 i$ & $12.6743 -2.7869 i$ & $12.2170 -4.6399 i$ & $13.1627 -0.9425 i$ & $12.9409 -2.8269 i$ & $12.4839 -4.7082 i$ \\
\hline
\end{tabular}
}
\end{center}
\caption{{\small The first three quasinormal frequencies for vector and tensor perturbation of the Schwarzschild black hole in $D=10$ dimensions.}}
\label{D10}
\end{table}

\section{The $D=9$ case.}

Here we give the details for the $D=9$ case, skipped in \cite{bcrst}. The substitution (\ref{odd}) leads to $2(D-3)=12$-term reccurence relation
\begin{equation}
\label{reccurenceD9}
\gamma^{(1)}_n a_{n+1} + \gamma^{(2)}_n a_{n} + ... + \gamma^{(12)}_n a_{n-10} = 0,
\end{equation}
with
\begin{eqnarray*}
\gamma^{(1)}_n &=& 216(1+ n)(ik- 3(1+ n))
\\
\gamma^{(2)}_n &=& - 9 (24 (5 + 12 n) ik + 40 k^2 - 252 + 147 s^2- 12 L(L+ 6)- 36 n (9 + 13 n))
\\
\gamma^{(3)}_n &=& 6 (12 (- 37 + 134 n) ik + 284 k^2 - 1323 + 1764 s^2- 36 L(L+ 6)+ 36 n (41 - 62 n))
\\
\gamma^{(4)}_n &=& - 9 (8 (- 289 + 292 n) ik + 452 k^2 - 4773 + 4312 s^2- 28 L(L+ 6)+ 12 n (515 - 253 n))
\\
\gamma^{(5)}_n &=& 4 (6 (- 2186 + 1279 n) ik + 1508 k^2 - 34290 + 21609 s^2- 36 L(L+ 6)+ 18 n (1870 - 547 n))
\\
\gamma^{(6)}_n &=& - (24 (- 3219 + 1324 n) ik + 6052 k^2 - 265050 + 130095 s^2- 36 L(L+ 6)+ 36 n (5579 - 1161 n))
\\
\gamma^{(7)}_n &=& 2 (12 (- 3109 + 986 n) ik + 2080 k^2 - 167877 + 69237 s^2+ 144 n (712 - 115 n))
\\
\gamma^{(8)}_n &=& - (48 (- 1007 + 260 n) ik + 1888 k^2 - 289215 + 105399 s^2+ 36 n (4097 - 541 n))
\\
\gamma^{(9)}_n &=& 8 (3 (- 849 + 185 n) ik + 64 k^2 - 21168 + 7056 s^2+ 9 n (1029 - 115 n))
\\
\gamma^{(0)}_n &=& - 2 (48 (- 53 + 10 n) ik + 32 k^2 - 32463 + 10143 s^2+ 18 n (691 - 67 n))
\\
\gamma^{(11)}_n &=& 6 (16 (- 6 + n) ik - 2463 + 735 s^2+ 24 n (35 - 3 n))
\\
\gamma^{(12)}_n &=& 9 (13 - 2 n+ 7 s) (13 - 2 n- 7 s)
\end{eqnarray*}
Gauss elimination \cite{Leaver2}, and insertion into (\ref{continuedfrac}) yields again quasinormal frequencies. The first three quasinormal frequencies for vector and tensor gravitational perturbation of the Schwarzschild black hole, for different values of $L$, are given in table \ref{D9}. 
\begin{table}
\begin{center}
{\scriptsize
\begin{tabular}{|c|c|c|c|c|c|c|}
\hline
$D=9$ & \multicolumn{3}{c|}{vector modes} & \multicolumn{3}{c|}{tensor modes}\\
\hline
L & $n=0$ & $n=1$ & $n=2$ & $n=0$ & $n=1$ & $n=2$ \\ 
\hline
 2 & $2.7928 -0.8542 i$ & $1.7792 -2.5965 i$ & $0.5699 -3.0121 i$ & $3.4488 -0.8601 i$ & $2.7548 -2.6116 i$ & $0.9903 -3.3373 i$ \\
 3 & $3.5389 -0.8278 i$ & $2.8438 -2.4808 i$ & $1.1129 -3.2806 i$ & $4.1342 -0.8541 i$ & $3.5853 -2.5825 i$ & $1.5262 -4.2850 i$ \\
 4 & $4.2863 -0.8180 i$ & $3.7573 -2.4494 i$ & $2.3546 -3.9155 i$ & $4.8200 -0.8506 i$ & $4.3624 -2.5661 i$ & $3.2201 -4.3333 i$ \\
 5 & $5.0260 -0.8161 i$ & $4.5953 -2.4469 i$ & $3.5813 -4.0450 i$ & $5.5061 -0.8483 i$ & $5.1123 -2.5559 i$ & $4.2023 -4.3007 i$ \\
 6 & $5.7578 -0.8171 i$ & $5.3913 -2.4524 i$ & $4.5724 -4.0828 i$ & $6.1926 -0.8469 i$ & $5.8463 -2.5491 i$ & $5.0773 -4.2796 i$ \\
 7 & $6.4826 -0.8191 i$ & $6.1617 -2.4597 i$ & $5.4646 -4.1045 i$ & $6.8792 -0.8458 i$ & $6.5699 -2.5443 i$ & $5.8994 -4.2651 i$ \\
 8 & $7.2018 -0.8214 i$ & $6.9152 -2.4669 i$ & $6.3037 -4.1198 i$ & $7.5660 -0.8451 i$ & $7.2863 -2.5409 i$ & $6.6898 -4.2547 i$ \\
 9 & $7.9165 -0.8236 i$ & $7.6569 -2.4735 i$ & $7.1097 -4.1317 i$ & $8.2529 -0.8445 i$ & $7.9975 -2.5383 i$ & $7.4592 -4.2470 i$ \\
10 & $8.6275 -0.8255 i$ & $8.3899 -2.4793 i$ & $7.8933 -4.1413 i$ & $8.9399 -0.8441 i$ & $8.7048 -2.5363 i$ & $8.2136 -4.2411 i$ \\
11 & $9.3354 -0.8273 i$ & $9.1161 -2.4844 i$ & $8.6607 -4.1492 i$ & $9.6269 -0.8438 i$ & $9.4091 -2.5348 i$ & $8.9571 -4.2365 i$ \\
12 & $10.0409 -0.8288 i$ & $9.8370 -2.4887 i$ & $9.4161 -4.1558 i$ & $10.3140 -0.8435 i$ & $10.1111 -2.5336 i$ & $9.6922 -4.2328 i$ \\
13 & $10.7442 -0.8301 i$ & $10.5537 -2.4925 i$ & $10.1620 -4.1614 i$ & $11.0011 -0.8433 i$ & $10.8112 -2.5325 i$ & $10.4207 -4.2298 i$ \\
14 & $11.4458 -0.8313 i$ & $11.2670 -2.4957 i$ & $10.9004 -4.1662 i$ & $11.6883 -0.8431 i$ & $11.5098 -2.5317 i$ & $11.1439 -4.2274 i$ \\
15 & $12.1460 -0.8323 i$ & $11.9774 -2.4986 i$ & $11.6328 -4.1703 i$ & $12.3755 -0.8430 i$ & $12.2070 -2.5310 i$ & $11.8628 -4.2254 i$ \\
\hline
\end{tabular}
}
\end{center}
\caption{{\small The first three quasinormal frequencies for vector and tensor gravitational perturbation of the Schwarzschild black hole in $D=9$ dimensions.}}
\label{D9}
\end{table}

\section{Summary.}
We have shown how to use Leaver's \cite{Leaver} method of continued fraction in the case of a number of regular singular points, which set lower bounds on the convergence radius of the series representation of a solution than irregular singular point. This prescription is general and together with Leaver's method makes a powerful tool in determination of quasinormal frequencies for wave equations and resonances in quantum mechanics.
 
\section*{Acknowledgments} I am greatly indebted to Piotr Bizo\'n for discussions, remarks and support in research. This work was supported by the Polish Ministry of Science Grant No. 1 P03B 012029.

\end{document}